# Switching LPV Approach for Analysis and Control of TCP-based Cyber-Physical Systems under DoS Attack

Soheila Barchinezhad, Vicenç Puig

*Abstract*—**Cyber-physical systems (CPSs) integrate controllers, sensors, actuators, and communication networks. Tight integration with communication networks makes CPSs vulnerable to cyberattacks. In this paper, we investigate the impact of denial of service (DoS) attack on the stability of cyber physical systems by considering the transmission control protocol (TCP) and extract a sufficient stability condition in linear matrix inequality (LMI) form. To this end, we model the TCP-CPS under DoS attack as a switching LPV time-delay system by using the Markov jump model. Then, we design parameter dependent stabilizing controller for CPS under DoS attack, by considering the network parameters. Finally, we prove efficiency and the feasibility of our findings through a well-known case study in the networked control systems literature.**

## 1. INTRODUCTION

Cyber physical systems (CPS) refer to integration of digital facilities ,encompassing communication network and computational capability, with physical devices and engineering systems [1]. The deep integration of physical space and communication networks, such as the Internet, exposes CPSs to the risk of cyberattacks. Consequently, safeguarding CPSs against such attacks has emerged as a prominent area of concern [2-5]. Cyber-physical security places a significant emphasis on ensuring the continuous availability and operability of plant, in addition to cyber security which is more concerned with data authenticity, integrity, and confidentiality. Cyberattacks in CPSs may affect some or all of these security services.

One of the primary categories of cyberattacks is Denial-of-Service (DoS) attacks, as referenced in [6, 7]. DoS attacks target the availability of a system by causing packet drops or significant delays in data streams. It is accomplished by consuming the communication or computational resources of closed-loop, such as the bandwidth, by sending a large number of data to the network such that system is unable to respond to normal service requests. These kind of attacks aimed to degrade the performance of the physical system or destabilize it. Hence, it is the motivation of many researches to study the stability of CPSs in the presence of DoS attacks.

There are a lot of studies in the stability analysis and stabilization of CPSs against DoS attacks, in the recent years [8-12]. For example, in [13], sufficient conditions on the duration and frequency of DoS attacks are proposed to guarantee stability of system. In [14], a packet-based stabilization method for CPS affected by DoS attacks, under assumption that the maximum number of DoS attacks is bounded, is designed. In the same way, [15] deals with the problems of data packet dropout and transmission delays induced by communication channel, and sufficient conditions on the stabilization of the closed-loop system are introduced. In [7, 16-18], stability analysis and/or controller design of CPSs under DoS attacks is studied using switching and Markov jump systems. All of the above papers consider the communication network as a black box which only has an unknown transmission delay and packet dropout. Despite the above papers, there are still many interesting problems, such as the network resource constraints and attack rate, which are important issue in examining the packet drops in CPS and is worth further exploration.

Stability and stabilization of CPS under different network parameters are not fully investigated. This motivates the present study. The amount of delay and drops induced by network and DoS attacks depend on the characteristics and protocols of communication network. One of the popular congestion control protocols used in the study of CPSs is transmission control protocol (TCP) [19-22], since CPSs need to have a reliable connection and TCP establishes that reliability. The combination of dynamic of TCP and the control system results on an augmented delay dependent time-varying system, in which the relation between dynamic of TCP and control system is established through a time varying delay. There are some works that studied the dynamic of TCP in CPS, in recent years. For instance, [23] proposes a robust controller for unmanned aerial vehicles (UAVs) by considering dynamic behavior of TCP. In [24], a method to control both the plant and the network using analytical TCP model as the network is presented. Also, [25] presents control systems under TCP communications as a system with both sensor-to-controller (SC) and controller-to-actuator (CA) delays and a queue management controller for network routers is designed to hold the network induced delays within their lower bounds.

In this paper, we investigate the stability and stabilization of CPSs in the presence of network induced delay and DoS attacks. The considered data transmission protocol in this study is TCP. By augmenting the control system with state-dependent delay differential equation of TCP, the closed-loop system is transformed to a nonlinear control system. We embedded the nonlinearities with linear parameter varying (LPV) approach, which can describe nonlinear systems by using the time-varying parameters. Because of the probabilistic nature of packet loss in network, this time delay LPV system returns to a stochastic switched one. Therefore, stability of CPSs in the presence of DoS attacks returns to the stability analysis of stochastic switched delay dependent LPV system that has both normal and attacked modes and switches randomly between these two modes. To implement the switching caused by switching DoS attack, we use Markovian jump system (MJS) which is a class of systems that can describe plants subject to random abrupt variations. In MJS, the switching between modes is according to a transition rate matrix, which in our model this matrix is affected by DoS attack rate and constraints of the network resources. There are many researches on stability analysis and stabilization of delay dependent LPV systems [26-29], but switched delay dependent LPV systems with parameter varying subsystems, have a few results. [30] investigate the switched LPV systems with time-dependent transition rate and LTI subsystems. [31] studied the design of stabilizing controller for switched LPV systems with Markovian jump parameters, with unknown or estimated transition rate. Also, [32] design controller for switched LPV systems by considering switching with average dwell time. As a further work, is required to study MJSs with delay dependent LPV subsystems and uncertain transition rates. We propose a sufficient condition for the stability of this class of systems using the parameter varying Lyapunov-Krasovskii and polytopic approach. Then, we design a stabilizing controller to guarantees the stability of system under DoS attacks. In a nutshell, the main contribution of paper is summarized as follows:

- We model the CPS under switching DoS attack by incorporating network parameters and dynamics as a delay dependent parameter varying Markov jump system (LPV-MJS).

- The parameter varying transition rate matrix of LPV-MJS is returned to a time-invariant one by estimation according to the dynamic of drops.

- We propose sufficient conditions for the stability of delay dependent LPV-MJS in LMIs using parameter-dependent Lyapunov function. To the best of authors' knowledge, no results have been proposed for the delay dependent case with parameter dependent subsystems.

- We solve the problem of infinite number of parameter-varying LMIs for delay dependent LPV-MJS using the convex polytopic structure.

- Then, we design a parameter-dependent stabilizing controller to guarantee stochastic stability of the closed-loop CPS under DoS attack.

- Finally, a set of simulations of our theoretical findings in Matlab/Simulink were conducted, which determine the tolerable DoS attack rate for an example plant.

The organization of this paper is described in the following. In Section II, the description and modeling of CPS in the presence of TCP as the network protocol are presented. The system is modelled using a continue-time parameter varying state space representation, with a particular emphasis on the analysis of switching DoS attacks. Section III introduces LMIs that provide a sufficient stability and stabilization condition for the presented system. In Section IV, the simulation results conducted on a sample CPS are reported and discussed. Finally, the concluding remarks are given in Section V.

**Notation:** The standard notations used in this paper are as follows. For the matrix $M$, $M^T$ denote its transpose, and $M > 0$ (resp. $M < 0$) means that $M$ is positive definite (resp. negative definite). $diag\{.\}$ signifies a diagonal matrix with its elements denoted by '$\{.\}$'. $\sup\{.\}$ represents the supremum of a set. The symbol '*' within a matrix denotes the symmetric element in block matrices.

## 2. PROBLEM SETUP

In this section, we present a mathematical model to describe the dynamics of a CPS subject to DoS attacks. We consider a linear control system in which controller and plant are connected through a network.

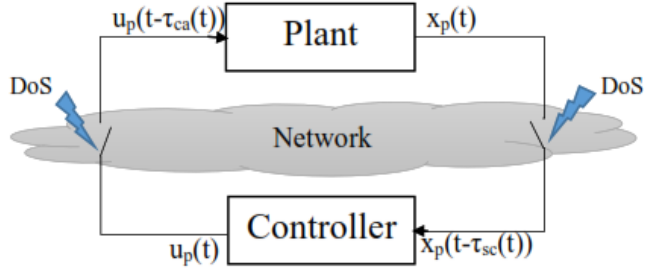

Fig. 1. Closed-loop system

### 2.1 Control System Dynamics

In this study, we use a linear time-invariant (LTI) control system model with state-feedback controller and network induced delay presented in Fig. 1 and following equation,

$$\begin{cases} \dot{x}_p(t) = A_p x_p(t) + B_p \overline{u}_p(t) \\ \overline{u}_p(t) = -K_p x_p(t - \tau(t)) \\ x_p(s) = \varphi_p(s), \quad s \in [-h, 0] \end{cases} \tag{1}$$

where $x_p(t)$ and $\overline{u}_p(t)$ are the system state and the delayed control input, respectively. The matrices $A_p$ and $B_p$ denote the state and input matrices. The function $\varphi_p$ represents the functional initial conditions of system. The control input function is considered as a state feedback gain $K_p$ to generate the control signals. Data transmission between the controller and the plant is carried out using the TCP and $\tau(t)$ represents $\tau_{sc}(t) + \tau_{ca}(t)$.

### 2.1 TCP modeling

The behavior of TCP was modeled with differential equations in [33-35]. These equations capture the dynamics of TCP's sending window and packet drops. Consider a network topology where TCP flows from different sources and different sessions traverse a shared bottleneck router, the behavior of the window size and bottleneck router queue size can be described by the following equations [33, 34],

$$\begin{cases} \dot{w}(t) = \dfrac{1}{\tau(t)} + (-\dfrac{w(t)}{2} \dfrac{w(t-\tau(t))}{\tau(t-\tau(t))} p_r(t-\tau(t))) \\ \dot{q}(t) = -C + \dfrac{N}{\tau(t)} w(t) + R_{DoS} \end{cases}$$

(2)

where $N$ is the number of eligible TCP sessions utilizing the shared bottleneck, $w(t) \in [1, w_m]$ denotes the sender window size (packets), $q(t) \in [0, q_m]$ represents the average queue length (packets), $R_{DoS}$ stands for the attacker data rate (packets/s), $p_r$ is the probability of data packet marking/dropping, and $\tau(t) = T_p + \dfrac{q(t)}{C}$ is the network delay (s), with $C$ the capacity of bottleneck link (packets/s), and $T_p$ the propagation delay. Additionally, $w_m$ and $q_m$ represent known maximum window size and queue buffer length, respectively.

To model the drop probability function mentioned in (2), we consider the RED model which is the most common drop function in Internet routers and is as [36],

$$p_r(t) = \begin{cases} 0, & if \quad 0 \leq q(t) < \min_{th} \\ \dfrac{q(t) - \min_{th}}{\max_{th} - \min_{th}} \max_p, & if \quad \min_{th} \leq q(t) \leq \max_{th} \\ 1, & if \quad q(t) > \max_{th} \end{cases}$$

(3)

with adjustable parameters $\min_{th}$, $\max_{th}$ and $\max_p$. There is a discontinuity at $\max_{th}$ which can be solved through smooth functions as follows

$$p_r(t) \approx (1 - f_{p2})(f_{p1} \dfrac{q(t) - \min_{th}}{\max_{th} - \min_{th}} \max_p) + f_{p2}, \qquad f_{p1} = \dfrac{1}{1 + e^{-b(q(t) - \min_{th})}}, \ f_{p2} = \dfrac{1}{1 + e^{-b(q(t) - \max_{th})}}$$

(4)

where $f_{p1}$ and $f_{p2}$ are smoothing functions, along with a positive integer value $b$.

The model given in (2) with considering (4), is a non-linear differential equation. Rewriting (2) and (4) in LPV state space yields to

$$\begin{cases} \dot{x}_n(t) = A_n(\rho(t)) x_n(t) + A_d(\rho(t)) x_n(t - \tau(t)) \\ x_n(s) = \varphi_n(s), \quad s \in [-h, 0] \end{cases}$$

(5)

with vector of states $x_n(t) = [q(t), w(t)]^T$, network time dependent delay $\tau(t)$, function of initial condition $\varphi_n(.)$, vector of varying parameters $\rho(t) = [\rho_1(t), \rho_2(t), \rho_3(t)]$, and $A_n(\rho(t)) = \begin{bmatrix} 0 & \rho_1(t) \\ 0 & \rho_2(t) \end{bmatrix}_{2 \times 2}$, $A_d(\rho(t)) = \begin{bmatrix} 0 & 0 \\ 0 & \rho_3(t) \end{bmatrix}_{2 \times 2}$,

where

$$\rho_1(t) = \frac{-C}{w(t)} + \frac{CN}{q(t) + CT_p} + \frac{R_{DoS}}{w(t)},$$

$$\rho_2(t) = \frac{C}{w(t)q(t) + w(t)CT_P},$$

$$\rho_3(t) = \frac{-Cw(t)}{2q(t-\tau(t)) + 2CT_p}((1-f_{p2})(f_{p1}\frac{q(t-\tau(t)) - \min_{th}}{\max_{th} - \min_{th}}\max_p) + f_{p2})$$

### 2.2 Augmented TCP-CPS Dynamics

By augmenting the control system model (1) that considers the TCP protocol model (5), it yields:

$$\begin{cases} \dot{x}(t) = A(\rho(t))x(t) + A_h(\rho(t))x(t-\tau(t)) \\ x(s) = \varphi(s), \quad s \in [-h, 0] \end{cases} \tag{6}$$

where $x(t) = [x_p(t), x_n(t)]^T$ is the augmented state, $\rho(t)$ is the vector of varying parameters, and matrices are

$$A(\rho(t)) = \begin{bmatrix} A_p & 0 \\ 0 & A_n(\rho(t)) \end{bmatrix}, A_h(\rho(t)) = \begin{bmatrix} -B_p K_p & 0 \\ 0 & A_d(\rho(t)) \end{bmatrix}.$$

The presented model for CPS in (6) is a LPV system with state dependent delay. Now, the subsequent part will focus on the modeling of a switched DoS for the described CPS.

### 2.3 System Dynamics under switching DoS Attack

Due to the stochastic nature of DoS attacks, a CPS can exhibit behavior similar to that of a switching system with multiple modes. A switching signal $r(t) \in \{0,1\}$ determines the status of the packet delivery. Case $r(t) = 0$ indicates that system is in normal mode and the plant receives the original data sent by the controller, albeit with delay. On the other hand, $r(t) = 1$ shows the sent data by the controller has been lost and the plant uses the input value $0$ instead. Consequently, we model the controller of system (1) as

$$K_p = \begin{cases} K_c & r(t) = 0 \\ 0 & r(t) = 1 \end{cases} \tag{7}$$

where $K_c$ is the gain matrix of control system. Therefore, switching model of LPV system (6) can be expressed as follows:

$$\begin{cases} \dot{x}(t) = A(r(t), \rho(t))x(t) + A_h(r(t), \rho(t))x(t-\tau(t)) \\ x(s) = \varphi(s), \quad s \in [-h, 0] \end{cases} \tag{8}$$

The switching system discussed in this paper is characterized as a continuous-time MJS. The process $r(t), t > 0$ involved in the system is a Markovian stochastic process taking values in a finite set $S = \{1, \ldots, M\}$ with transition probabilities

$$\Pr\{r(t + \Delta t) = j | r(t) = i\} = \begin{cases} \pi_{ij} \Delta t + o(\Delta t), & i \neq j \\ 1 + \pi_{ii} \Delta t + o(\Delta t), & i = j \end{cases} \qquad (9)$$

where $\pi_{ij}$ is the transition rate from mode $i$ at sample time $t$ to mode $j$ at sample time $t + \Delta t$ with $\Delta t > 0$, $\lim_{\Delta t \to 0} \frac{o(\Delta t)}{\Delta t} = 0$ and

$\pi_{ii} = -\sum_{j=1, j \neq i}^{M} \pi_{ij}$. Let $\Pi = [\pi_{ij}]$, then by considering the probability of drop ($p_r(t)$),

$$\Pi = \begin{bmatrix} -p_r(t) & p_r(t) \\ 1 - p_r(t) & -(1 - p_r(t)) \end{bmatrix} \qquad (10)$$

Based on Equation (4), $p_r(t)$ depends on bottleneck queue length ($q(t)$) which can take values from the set $[0, 1, 2, \ldots, q_m]$. As a result, system (8) can be classified as delay dependent LPV-MJS with $M = 2(q_m + 1)$ modes. The first $(q_m + 1)$ modes indicate the un-attacked mode, where the queue size ranges from $0$ to $\mathrm{q_m}$. In the same manner, the last $(q_m + 1)$ modes correspond to attacked mode. Consequently, the transition rate matrix (10) can be transformed into

$$\Pi = \begin{bmatrix} \hat{\pi}_{11} + \Delta_{11} & \cdots & \hat{\pi}_{1(q_m+1)} + \Delta_{1(q_m+1)} & \hat{\pi}_{1(q_m+2)} + \Delta_{1(q_m+2)} & \cdots & \hat{\pi}_{1(2q_m+2)} + \Delta_{1(2q_m+2)} \\ \hat{\pi}_{21} + \Delta_{21} & \cdots & \hat{\pi}_{2(q_m+1)} + \Delta_{2(q_m+1)} & \hat{\pi}_{2(q_m+2)} + \Delta_{2(q_m+2)} & \cdots & \hat{\pi}_{2(2q_m+2)} + \Delta_{2(2q_m+2)} \\ \vdots & & \vdots & & & \\ \hat{\pi}_{(q_m+1)1} + \Delta_{(q_m+1)2} & \cdots & \hat{\pi}_{(q_m+1)(q_m+1)} + \Delta_{(q_m+1)(q_m+1)} & \hat{\pi}_{(q_m+1)(q_m+2)} + \Delta_{(q_m+1)(q_m+2)} & \hat{\pi}_{(q_m+1)(2q_m+2)} + \Delta_{(q_m+1)(2q_m+2)} \\ \hat{\pi}_{(q_m+2)1} + \Delta_{(q_m+2)2} & \cdots & \hat{\pi}_{(q_m+2)(q_m+1)} + \Delta_{(q_m+2)(q_m+1)} & \hat{\pi}_{(q_m+1)(q_m+2)} + \Delta_{(q_m+1)(q_m+2)} & \hat{\pi}_{(q_m+2)(2q_m+2)} + \Delta_{(q_m+2)(2q_m+2)} \\ \vdots & & \vdots & & & \vdots \\ \hat{\pi}_{(2q_m+1)1} + \Delta_{(2q_m+1)2} & \cdots & \hat{\pi}_{(2q_m+2)(q_m+1)} + \Delta_{(2q_m+1)(q_m+1)} & \hat{\pi}_{(q_m+1)(q_m+2)} + \Delta_{(2q_m+1)(q_m+2)} & \hat{\pi}_{(2q_m+1)(2q_m+2)} + \Delta_{(2q_m+1)(2q_m+2)} \\ \hat{\pi}_{(2q_m+2)1} + \Delta_{(2q_m+2)2} & \cdots & \hat{\pi}_{(2q_m+2)(q_m+1)} + \Delta_{(2q_m+2)(q_m+1)} & \hat{\pi}_{(q_m+1)(q_m+2)} + \Delta_{(2q_m+2)(q_m+2)} & \hat{\pi}_{(2q_m+2)(2q_m+2)} + \Delta_{(2q_m+2)(2q_m+2)} \end{bmatrix} \qquad (11)$$

with

$$\hat{\pi}_{ij} = \begin{cases} 1 - p_r(q(t) = j-1) & if \ i \neq j, (j \leq q_m + 1) \\ p_r(q(t) = j - (q_m + 1) - 1) & if \ i \neq j, (j > q_m + 1) \\ -\sum_{j=1, j \neq i}^{M} \hat{\pi}_{ij} & if \ i = j \end{cases} \qquad (12)$$

and $\Delta_{ij} \in [-\delta_{ij}, \delta_{ij}]$, where $\delta_{ij} > 0$ represents the known estimated error of transition rate, which can be calculated with (4).

$$\delta_{ij} = \begin{cases} p_r(q(t) = j) - p_r(q(t) = j-1) & if \ i \neq j, (j \leq q_m + 1) \\ p_r(q(t) = j - (q_m + 1)) - p_r(q(t) = j - (q_m + 1) - 1) & if \ i \neq j, (j > q_m + 1) \\ -\sum_{j=1, j \neq i}^{M} \delta_{ij} & if \ i = j \end{cases} \qquad (13)$$

**Assumption 1:** We make the following assumptions:

$$0 < \tau(t) \le \bar{h} \tag{14}$$

$$0 \le \dot{\tau}(t) < \mu \tag{15}$$

Here, $\bar{h}$ represents the upper bounds of $\tau(t)$ that guarantees the stability of system (8) and $\mu = \sup_{w(t),q(t)} \{\dot{\tau}(t)\}$.

Based on these assumptions, we proceed to analyze the stochastic stability of the delay-dependent LPV-MJS (8) with the Markovian jump model in the subsequent section.

### 3. MAIN RESULTS

The primary focus of this section is to analyze the stability of the switching LPV system in the presence of switches triggered by DoS attacks. We aim to provide sufficient conditions, expressed in the form of LMIs, to ensure the stochastic stability of the system (8) when subjected to DoS attacks modeled in (7). By analyzing these conditions, we can ascertain the stability of the system and its resilience against DoS attacks.

*3.1 Stability Analysis*

**Theorem 1**. Consider the switched time-delay LPV system (8) with Markovian jump parameter $r(t)$ and the transition rate matrix (11). It is stochastically stable for a given $\bar{h} > 0$, if there exist symmetric matrix functions $P_i > 0$ and symmetric matrices $R > 0$ and $Q > 0$, such that

$$\Xi(\rho(t)) = \begin{bmatrix} \Xi_{11} & \Xi_{12} & \bar{h} A_i^T(\rho(t))R \\ * & \Xi_{22} & \bar{h} A_{hi}^T(\rho(t))R \\ * & * & -R \end{bmatrix} \tag{16}$$

with

$$\Xi_{11} = A_i^T(\rho(t))P_i(\rho(t)) + P_i(\rho(t))A_i(\rho(t)) + \left( \frac{dP_i(\rho(t))}{dt} \right) + \sum_{j=1}^{N}(\pi_{ij} + \delta_{ij})P_j(\rho(t)) + Q - R$$

$$\Xi_{12} = P_i(\rho(t))A_{hi}(\rho(t)) + R$$

$$\Xi_{22} = -(1 - \sup_{w(t),q(t)}\{\dot{\tau}(t)\})Q - R$$

holds for all switching modes $i$.

**Proof**: To analyze the stability of the switching system (8), we utilize a parameter-dependent Lyapunov function, inspired by the works of [26, 27]. Let us define this Lyapunov function as follows:

$$V(x(t), r(t), \rho(t)) = x^T(t)P(r(t), \rho(t))x(t) + \int_{t-\tau(t)}^{t} x^T(s)Qx(s)ds + \bar{h}\int_{-\bar{h}}^{0}\int_{t+\theta}^{t}\dot{x}^T(s)R\dot{x}(s)dsd\theta \tag{17}$$

Then, we obtain

$$\dot{V}(x(t), \rho(t)) = \dot{x}^T(t)P_i(\rho(t))x(t) + x^T(t)P_i(\rho(t))\dot{x}^T(t) + x^T(t)\left(\frac{dP_i(\rho(t))}{dt}\right)x(t) + x^T(t)\sum_{j=1}^{N}\pi_{ij}P_j(\rho(t))x(t)$$
$$+ x^T(t)Qx(t) - (1-\dot{\tau}(t))x^T(t-\tau(t))Qx(t-\tau(t)) + \bar{h}^2\dot{x}^T(t)R\dot{x}(t) - \bar{h}\int_{t-\bar{h}}^{t}\dot{x}^T(\eta)R\dot{x}(\eta)d\eta \tag{18}$$

According to (15) $\dot{\tau}(t) \leq \mu$, thus we have $-(1-\dot{\tau}(t)) \leq -(1 - \sup_{w(t), q(t)}\{\dot{\tau}(t)\})$. On the other hand, since $\tau(t) \leq \bar{h}$, it results

$$-\bar{h}\int_{t-\bar{h}}^{t}\dot{x}(\theta)^T R\dot{x}(\theta)d\theta \leq -\bar{h}\int_{t-\tau(t)}^{t}\dot{x}(\theta)^T R\dot{x}(\theta)d\theta \tag{19}$$

Using Jensen's inequality, we bound the right hand of (19),

$$-\int_{t-\tau(t)}^{t}\dot{x}(\theta)^T R\dot{x}(\theta)d\theta \leq -\frac{1}{h}(\int_{t-\tau(t)}^{t}\dot{x}(\theta)^T d\theta)^T R(\int_{t-\tau(t)}^{t}\dot{x}(\theta)d\theta) \tag{20}$$

According to Leibniz-Newton model transformation, $\int_{t-\tau(t)}^{t}\dot{x}(\theta)^T d\theta = x(t) - x(t-\tau(t))$, the following equation holds.

$$-\bar{h}\int_{t-\tau(t)}^{t}\dot{x}(\theta)^T R\dot{x}(\theta)d\theta \leq -[x(t) - x(t-\tau(t))]^T R[x(t) - x(t-\tau(t))] \tag{21}$$

Therefore,

$$\dot{V}(x(t), \rho(t)) \leq \dot{x}^T(t)P_i(\rho(t))x(t) + x^T(t)P_i(\rho(t))\dot{x}^T(t) + x^T(t)\left(\frac{dP_i(\rho(t))}{dt}\right)x(t) + x^T(t)\sum_{j=1}^{N}(\hat{\pi}_{ij} + \delta_{ij})P_j(\rho(t))x(t)$$
$$+ x^T(t)Qx(t) - (1 - \sup_{w(t), q(t)}\{\dot{\tau}(t)\})x^T(t-\tau(t))Qx(t-\tau(t)) + \bar{h}^2\dot{x}^T(t)R\dot{x}(t) - [x(t)-x(t-\tau(t))]^T R[x(t)-x(t-\tau(t))] \tag{22}$$

Thus, with replacing (8) in (22), it yields

$$\dot{V}(x(t), \rho(t)) \leq X^T(t)\Phi(\rho(t))X(t) \tag{23}$$

where

$$X^T(t) = [x^T(t), x^T(t-\tau(t))] \text{ and } \Phi(\rho(t)) = \begin{bmatrix} \Phi_{11} & \Phi_{12} \\ * & \Phi_{22} \end{bmatrix}$$

with

$$\Phi_{11} = A_i^T(\rho(t))P_i(\rho(t)) + P_i(\rho(t))A_i(\rho(t)) + \left(\frac{dP_i(\rho(t))}{dt}\right) + \sum_{j=1}^{N}(\hat{\pi}_{ij} + \delta_{ij})P_j(\rho(t)) + Q + \bar{h}^2[A_i^T(\rho(t))RA_i(\rho(t))] - R$$

$$\Phi_{12} = P_i(\rho(t))A_{hi}(\rho(t)) + \bar{h}^2[A_i^T(\rho(t))RA_{hi}(\rho(t))] + R$$

$$\Phi_{22} = -(1 - \sup_{w(t),q(t)}\{\dot{\tau}(t)\})Q + \overline{h}^2[A_{hi}{}^T(\rho(t))RA_{hi}(\rho(t))] - R$$

By using the Schur complement formula to matrix inequality (16), it yields $\Phi(\rho(t)) < 0$ which means $\dot{V}(\rho(t)) < 0$. However, it is important to note that the matrix inequality presented in Theorem 1 represents an infinite-dimensional condition, which implies satisfying an infinite number of conditions. To address this issue, a commonly employed approach is to transform the parameter-dependent LMI condition into a finite set of LMIs using the polytopic framework. The polytopic system of (8) can be formulated in general form as:

$$\begin{cases} \dot{x}(t) = \sum_{k=1}^{M} \lambda_k(\rho(t))(A^k x(t) + A_h{}^k x(t - \tau(t))) \\ x(s) = \varphi(s), \quad s \in [-h, 0] \end{cases} \quad (24)$$

with coefficients satisfying $\lambda_k(\rho(t)) \geq 0$, $\sum_{k=1}^{M} \lambda_k(\rho(t)) = 1$, and matrices $P_i(\rho(t)), Q$ and $R$ are also assumed to be as

$P_i(\rho(t)) = \sum_{k=1}^{M} \lambda_k(\rho(t))P_i^k$, $Q = \sum_{k=1}^{M} \lambda_k(\rho(t))Q^k$ and $R = \sum_{i=1}^{M} \lambda_k(\rho(t))R^k$, then the following relaxation result will be provided.

**Theorem 2.** For a given scalars $\overline{h} > 0$, MJS (8) is stochastically stable, if there exist symmetric matrices $P_i^k > 0$, $R^k > 0$ and $Q^k > 0$, for all polytope vertices $k$ and all switching modes $i$, such that the following LMI holds:

$$\Xi^k = \begin{bmatrix} \Xi_{11} & \Xi_{12} & \overline{h}(A_i^k)^T R^k \\ * & \Xi_{22} & \overline{h}(A_{hi}^k)^T R^k \\ * & * & -R^k \end{bmatrix} \quad (25)$$

with

$$\Xi_{11} = (A_i^k)^T P_i^k + P_i^k A_i^k + \left(\frac{dP_i(\rho(t))}{dt}\right) + \sum_{j=1}^{N}(\hat{\pi}_{ij} + \delta_{ij})P_j^k + Q^k - R^k$$

$$\Xi_{12} = P_i^k A_{hi}^k + R^k$$

$$\Xi_{22} = -(1 - \sup_{w(t),q(t)}\{\dot{\tau}(t)\})Q^k - R^k$$

**Remark 1.** The parameters of vector $\rho(t) = [\rho_1(t), \rho_2(t), \rho_3(t)]$ are bounded. These parameters can be measured or estimated in real-time, as the variable $w(t)$ is known to the sender and can be adjusted based on packet drops. Additionally, the parameter $q(t)$ can be estimated by considering the drop probability as described in equation (4). Thus, the polytopic model of system (8) is expressed as a function of the extreme values of parameters which are the vertices of the bounding box.

### 3.2 Design of parameter-dependent controller

In the following, a condition for the stabilization of the delay dependent LPV system (8) is derived. We find parameter-dependent stabilizing state-feedback controller using LMIs.

Consider the following form of control law:

$$\bar{u}(t) = -K(\rho(t))x(t - \tau(t))$$ (26)

System of (6) can be rewritten as:

$$\begin{cases} \dot{x}(t) = A(\rho(t))x(t) + A_{hn}(\rho(t))x(t - \tau(t)) - BK(\rho(t))x(t - \tau(t)) \\ x(s) = \varphi(s), \quad s \in [-h, 0] \end{cases}$$ (27)

in which

$$A(\rho(t)) = \begin{bmatrix} A_p & 0 \\ 0 & A_n(\rho(t)) \end{bmatrix}, A_{hn}(\rho(t)) = \begin{bmatrix} 0 & 0 \\ 0 & A_d(\rho(t)) \end{bmatrix}, B = \begin{bmatrix} B_p \\ 0 \end{bmatrix}.$$

By considering switching DoS attack as (7), switching model of LPV system (27) is as:

$$\begin{cases} \dot{x}(t) = A(r(t), \rho(t))x(t) + A_{hn}(r(t), \rho(t))x(t - \tau(t)) - BK(r(t), \rho(t))x(t - \tau(t)) \\ x(s) = \varphi(s), \quad s \in [-h, 0] \end{cases}$$ (28)

**Theorem 3.** Consider the switched LPV system (28) with Markovian jump parameters in (10) and (11). There exists a parameter-dependent state-feedback controller (26) such that the closed-loop system (28) is stochastically stable for a given $\bar{h} > 0$, if there exist matrix functions $P_i > 0$, $Y_i$ and matrices $R > 0$ and $Q > 0$, such that

$$\Gamma(\rho(t)) = \begin{bmatrix} \Gamma_{11} & \Gamma_{12} & \bar{h}P_i(\rho(t))A_i^T(\rho(t)) \\ * & \Gamma_{22} & \Gamma_{23} \\ * & * & -R \end{bmatrix}$$ (29)

with

$$\Gamma_{11} = P_i(\rho(t))A_i^T(\rho(t)) + A_i(\rho(t))P_i(\rho(t)) + \left(\frac{dP_i(\rho(t))}{dt}\right) + \sum_{j=1}^{N}(\hat{\pi}_{ij} + \delta_{ij})P_j(\rho(t)) + Q - R$$

$$\Gamma_{12} = A_{hni}(\rho(t))P_i(\rho(t)) - BY(\rho(t)) + R$$

$$\Gamma_{22} = -(1 - \sup_{w(t),q(t)}\{\dot{\tau}(t)\})Q - R$$

$$\Gamma_{23} = \bar{h}P_i(\rho(t))A_{hn}^T(\rho(t)) - \bar{h}Y^T(\rho(t)))B^T$$

holds for all switching modes $i$. Then,

$$K_i(\rho(t)) = Y_i(\rho(t))P_i(\rho(t))^{-1}.$$ (30)

**Proof.** Substitute the closed-loop system (28) into inequality (16). Then perform a congruence transformation with respect to matrix $diag\left\{P_i^{-1}(\rho(t)), P_i^{-1}(\rho(t)), R^{-1}\right\}$, and define variable changes

$$P_i(\rho(t)) \leftarrow P_i^{-1}(\rho(t)), R \leftarrow P_i^{-1}(\rho(t))RP_i^{-1}(\rho(t)), Q \leftarrow P_i^{-1}(\rho(t))QP_i^{-1}(\rho(t)), Y(\rho(t)) \leftarrow K(\rho(t))P_i(\rho(t))$$

The proof is completed.

If the polytopic system of (28) is formulated as:

$$\begin{cases} \dot{x}(t) = \sum_{k=1}^{M} \lambda_k(\rho(t))(A^k x(t) + A_{hn}{}^k x(t-\tau(t)) - BK^k x(t-\tau(t))) \\ x(s) = \varphi(s), \quad s \in [-h, 0] \end{cases} \tag{31}$$

with coefficients that satisfy $\lambda_k(\rho(t)) \geq 0$, $\sum_{k=1}^{M} \lambda_k(\rho(t)) = 1$, and matrices of (29) is also assumed to be as

$P_i(\rho(t)) = \sum_{k=1}^{M} \lambda_k(\rho(t)) P_i^k(\rho(t))$, $Q = \sum_{k=1}^{M} \lambda_k(\rho(t)) Q^k$ and $R = \sum_{i=1}^{M} \lambda_k(\rho(t)) R^k$, then the following result will be provided.

**Theorem 4.** For given scalars $\bar{h} > 0$ and $\mu \geq 0$, MJS (8) is stochastically stabilizable, if there exist symmetric matrices $P_i^k > 0$, $R^k > 0$, $Q^k > 0$ and $Y_i^k$, for all $k$ and all switching modes $i$, such that the following LMI holds:

$$\Gamma(\rho(t)) = \begin{bmatrix} \Gamma_{11} & \Gamma_{12} & \bar{h} P_i^k (A_i^k)^T \\ * & \Gamma_{22} & \Gamma_{23} \\ * & * & -R^k \end{bmatrix} \tag{32}$$

where

$$\Gamma_{11} = P_i^k (A_i^k)^T + A_i^k P_i^k + \sum_{j=1}^{N}(\hat{\pi}_{ij} + \delta_{ij}) P_j^k + \left(\frac{dP_i(\rho(t))}{dt}\right) + Q^k - R^k$$

$$\Gamma_{12} = A_{hni}{}^k P_i^k - BY_i^k + R^k$$

$$\Gamma_{22} = -(1 - \sup_{w(t),q(t)}\{\dot{\tau}(t)\})Q^k - R^k$$

$$\Gamma_{23} = \bar{h} P_i^k (A_{hni}^k)^T - \bar{h}(Y_i^k)^T B^T$$

Then,

$$K_i^k = (Y_i^k)^{-1} P_i^k . \tag{33}$$

## 4 APPLICATION EXAMPLE

In order to demonstrate the practical application and validation of the theoretical results presented earlier, we will provide an example based on existing literature [24, 37-41], where the plant is an unmanned ground vehicle (UGV) with parameters as follows:

$$\begin{bmatrix} \dot{x} \\ \dot{v} \end{bmatrix} = \begin{bmatrix} 0 & 1 \\ 0 & -0.1 \end{bmatrix} \begin{bmatrix} x \\ v \end{bmatrix} + \begin{bmatrix} 0 \\ 0.1 \end{bmatrix} F \tag{34}$$

where $x$, $v$ and $F$ are the position, velocity, and input force of the system, respectively. To control the states of system under normal conditions, the state feedback controller $K_c = [3.75, 11.5]$ is introduced in the mentioned references.

In our simulations, we utilized network parameters as outlined in Table 1. The table provides a comprehensive overview of the specific network parameters employed during the simulations for this particular application.

Table 1. Network parameters used in simulation

| | |
|---|---|
| **Number of TCP sessions (N)** | 100 |
| **TCP window size (w)** | [1-$w_m$] |
| **Maximum window size ($w_m$)** | 20 packet |
| **Queue length (q)** | [1-$q_m$] |
| **Maximum Queue length ($q_m$)** | 200 or 300 packet |
| **Propagation delay ($T_p$)** | 0.2 s |
| **Bottleneck link capacity (C)** | 5000– 8000 packet/s |

*4.1 Stability Analysis*

In this section, our objective is to investigate the stability of the system (34) under DoS attack. Specifically, we aim to determine the maximum tolerable DoS attack rate that the stability of system under that can be guaranteed. To achieve this, we explore the maximum allowable transmission delay bound that will not destabilize the system. It is important to note that DoS attacks can disrupt the system by causing packet drops and introducing significant delays.

Our findings reveal that the maximum allowable delay bound, which guarantees system stability, is influenced by network properties. These properties include the output capacity of the bottleneck link ($C$) and the capacity of the bottleneck buffer ($q_m$). For instance, our analysis using Theorem 1 reveals that the maximum allowable delay, for C=5000, $q_m = 300$ and the case that there is no DoS attack ($R_{DoS} = 0$), is 8.24, while the tolerable delay for C=5000 and $q_m = 200$ is 7.6. These results are more reasonable in comparison to previous works reported in Table 2, as our approach takes into account the effects of network parameters on the tolerable delay in a CPS. By considering these network properties, we are better equipped to provide a more accurate assessment of the maximum allowable delay that guarantees system stability in the presence of DoS attacks.

Table 2. Maximum allowable delay bound

| | |
|---|---|
| [37] | 0.0538 |
| [38] | 0.8695 |
| [39] | 0.9350 |
| [24] | 1.0029 |
| [41] | 1.7 |
| [40] | 26.3 |
| Theorem 1 with R_DoS=0, C=5000 and $q_m$=200 | **7.6** |
| Theorem 1 with R_DoS=0, C=5000 and $q_m$=300 | **8.24** |

To further validate our approach, we examined the applicability of our theorem for different attack rates $0, 50, 100, \ldots, 500$, while considering $C = 5000$ and $q_m = 200 \; or \; 300$. The results are presented in Fig. 2, showcasing the relationship between the attack rate and the allowable transmission delay. As the attack rate increases, the allowable transmission delay decreases. These findings also indicate that our approach can guarantee the stability of the system if the attack rate is less than or equal to 300 for $q_m = 200$. Furthermore, when the attack rate exceeds 350, the allowable delay becomes smaller than the maximum network delay $(\tau = T_p + \frac{q_m}{C})$, indicating that the stability of system cannot be guaranteed. To further explore the impact of network properties on the maximum tolerable DoS attack rate, we calculated the tolerable attack rate for various bottleneck output link capacities and reported the results in Table 3. The table demonstrates that as the bottleneck output link capacity increases, the maximum tolerable DoS attack rate also increases. For example, when C is 5000, the maximum tolerable DoS attack rate is 300, whereas for C=8000, the maximum tolerable DoS attack rate is 6400. These findings emphasize the significance of considering network properties when designing secure and stable CPS.

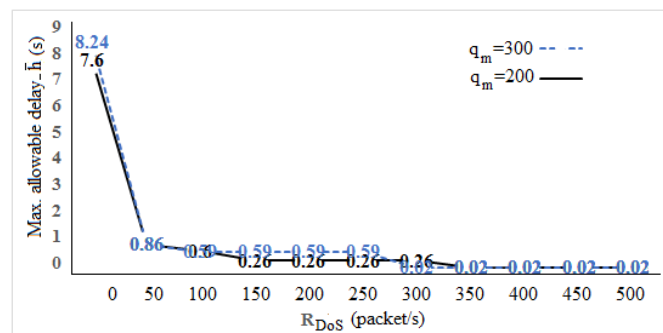

Fig. 2. Maximum allowable delay for CPS under different DoS attack rate

Table 3. Maximum tolerable RDoS ($q_m$=200)

| C | 5000 | 5500 | 6000 | 6500 | 7000 | 7500 | 8000 |
|---|---|---|---|---|---|---|---|
| $R_{DoS}$ | 300 | 1300 | 2350 | 3350 | 4400 | 5400 | 6400 |

### 4.2 Controller design

Guided by Theorem 4, we design the stabilization controller for system (34). Our objective is to evaluate the performance of this controller under various rates of DoS attacks. Figure 3 presents the results of these examinations, considering the network parameters specified in Table 1 with $q_m = 200$ and C=5000. The results demonstrate the effectiveness of the stabilization controller in the presence of DoS attacks. The findings demonstrate that the proposed controller is capable of maintaining system behavior stable even in the presence of high rates of DoS attacks. This indicates the robustness and resilience of the controller, as it successfully counteracts the destabilizing effects of DoS attacks on the CPS. These results are significant as they demonstrate the controller's ability to enhance the security and stability of CPS in the face of DoS attacks. Overall, the findings presented in this section suggest that the stabilization controller proposed in this paper can effectively address the issue of DoS attacks in CPS. This could have important implications for the development of secure and robust CPS.

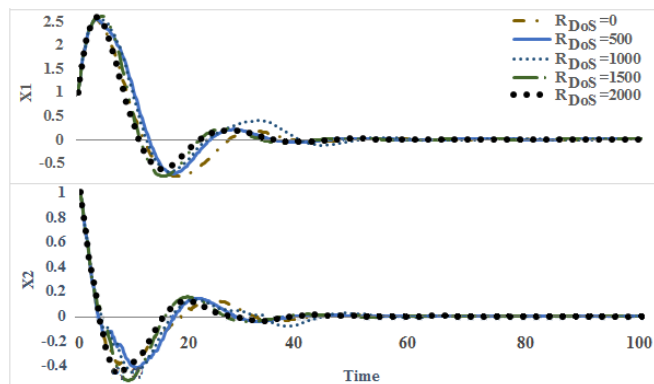

Fig. 3. State trajectory of the plant (34) with proposed stabilizing controller.

## 5 CONCLUSION

This paper focused on addressing the stability and stabilization of cyber physical systems (CPSs) in the presence of switched Denial-of-Service (DoS) attack. It emphasized the importance of considering network parameters in the modeling and analyzing the CPS under DoS attacks. The study specifically focused on the TCP protocol, known for its resilience in mitigating the impact of DoS attacks. By incorporating TCP dynamics into the CPS model, a nonlinear parameter-varying system was created. To account for the nonlinearities, the Linear Parameter-Varying (LPV) approach was employed. The research investigated the stability and control problem of the switched LPV system using a Markov jump model, where the transition rates were estimated based on network parameters. It introduced stability conditions in Linear Matrix Inequalities (LMIs) to find the tolerable delay and DoS attack rates under which the CPS stability is guaranteed. Then a stabilization controller is designed to mitigate the effect of DoS attacks. The proposed theoretical findings were evaluated through simulations using a control plant example, considering different values for network parameters, and the results validated the effectiveness of the approach.